\documentclass[12pt,preprint]{aastex}

\shorttitle{NGC1313 X-2 age}
\shortauthors{}

\title{Super-Eddington Accretion in the Ultraluminous X-ray Source NGC1313 X-2:
An Ephemeral Feast}

\author{Shan-Shan Weng\altaffilmark{1,2,3}, Shuang-Nan Zhang\altaffilmark{2,4,5},
Hai-Hui Zhao\altaffilmark{2}}

\email{wengss@ihep.ac.cn; zhangsn@ihep.ac.cn; zhaohh@ihep.ac.cn}

\altaffiltext{1}{Department of Physics, Xiangtan University, Xiangtan 411105,
China} \altaffiltext{2} {Key Laboratory of Particle Astrophysics, Institute of
High Energy Physics, Chinese Academy of Sciences, Beijing 100049, China}
\altaffiltext{3} {Sabanc\i~University, Faculty of Engineering and Natural
Sciences, Orhanl\i$-$ Tuzla, \.{I}stanbul 34956, Turkey} \altaffiltext{4}
{National Astronomical Observatories, Chinese Academy of Scienc es, Beijing,
100012, China}\altaffiltext{5} {Physics Department, University of Alabama in
Huntsville, Huntsville, AL 35899, USA}

\begin{document}
\begin{abstract}

We investigate the X-ray spectrum, variability and the surrounding ionized
bubble of NGC1313 X-2 to explore the physics of super-Eddington accretion.
Beyond the Eddington luminosity, the accretion disk of NGC1313 X-2 is truncated
at a large radius ($\sim$ 50 times of innermost stable circular orbit), and
displays the similar evolution track with both luminous Galactic black-hole and
neutron star X-ray binaries. In super-critical accretion, the speed of
radiatively driven outflows from the inner disk is mildly relativistic. Such
ultra-fast outflows would be over ionized and might produce weak Fe K
absorption lines, which may be detected by the coming X-ray mission {\it
Astro-H}. If the NGC1313 X-2 is a massive stellar X-ray binary, the high
luminosity indicates that an ephemeral feast is held in the source. That is,
the source must be accreting at a hyper-Eddington mass rate to give the
super-Eddington emission over $\sim 10^{4}-10^{5}$ yr. The expansion of the
surrounding bubble nebula with a velocity of $\sim$ 100 km~s$^{-1}$ might
indicate that it has existed over $\sim 10^{6}$ yr and is inflated by the
radiatively driven outflows from the transient with a duty cycle of activity of
$\sim$ a few percent. Alternatively, if the surrounding bubble nebula is
produced by line-driven winds, less energy is required than the radiatively
driven outflow scenario, and the radius of the Str\"{o}mgren radius agrees with
the nebula size. Our results are in favor of the line-driven winds scenario,
which can avoid the conflict between the short accretion age and the apparently
much longer bubble age inferred from the expansion velocity in the nebula.

\end{abstract}


\section{Introduction}
Hundreds of non-nuclear X-ray sources have been found in nearby galaxies with
isotropic luminosities $> 10^{39}$ erg/s, now called ultraluminous X-ray
sources (ULXs, Liu \& Mirabel 2005; Swartz et al. 2011). Recently, Feng \&
Soria (2011) reviewed the multiwavelength studies of ULXs, and summarized that
ULXs are a diverse population (see also Fender \& Belloni 2012). A handful of
ULXs contain normal stellar mass black holes (BHs), with a mass $M_{\rm BH}$
$\sim$ 5--15 $M_{\odot}$ (Weng et al. 2009; Middleton et al. 2012). A few
brightest ULXs are strong candidates of intermediate mass BHs (IMBHs), $M_{\rm
BH}$ $\sim$ $10^{2}-10^{4}$ $M_{\odot}$, e.g. ESO243-49 HLX-1 (Farrell et al.
2009; Webb et al. 2012) and M82 X-1 (Strohmayer \& Mushotzky 2003). The
majority of ULXs are thought to be massive stellar black hole X-ray binaries
(MXBs), $M_{\rm BH}$ $\sim$ 50--100 $M_{\odot}$ (Patruno \& Zampieri 2008).

The mass of a BH depends both on the initial mass and metallicity of stellar
progenitor (Heger et al. 2003; Zampieri \& Roberts 2009; Belczynski et al.
2010). The low metallicity environments in the early universe are suitable for
massive stars producing massive BHs. The role of MXBs in cosmology has
attracted increasing attention in recent years (e.g. Wheeler \& Johnson 2011;
Fragos et al 2013; Power et al. 2013). Mirabel et al. (2011) argued that
besides the ultraviolet (UV) radiation from massive stars, feedback from MXBs
is an additional, important source of heating and reionization of the Inter
Galactic Medium at the dawn of the universe. Justham \& Schawinski (2012)
suggested that the MXB feedback might significantly affect the galaxy
formation, especially the dwarf galaxies and the earliest epoch of galaxy
formation. If most ULXs are MXBs, they are a good sample for us to study all
these physical processes. The luminosities $\sim$ $10^{40}$ erg/s illustrates
super-Eddington accretion in this population of ULXs (Gladstone et al. 2009;
Weng et al. 2009).

However, due to various reasons, our understanding of super-Eddington accretion
flow is extremely rudimentary. There is a subclass of active galactic nuclei
(AGNs) that are candidates for super-Eddington accretion, namely, narrow line
Seyfert 1 galaxies (Collin \& Kawaguchi 2004; Middleton \& Done 2010). However,
such sources are far away from the earth, and their environments are relatively
complicated. Collecting a few bright Galactic sources, Weng \& Zhang (2011,
Paper I) found that BH and neutron star (NS) X-ray binaries (XRBs) shared the
same accretion disk evolution pattern when approaching Eddington luminosity
(see also Neilsen et al. 2011). Through the observed NS surface emission, in
Paper I we argued that the disk thickness H/R is less than 0.3-0.4, and the
significant outflow from inner disk edge exists at luminosity close to
Eddington luminosity.

NGC1313 X-2 is one of the best-studied ULXs, which has the richest observation
data at both X-ray and optical wavelengths (Liu et al. 2007; Pintore \&
Zampieri 2012). The source was first detected by {\it Einstein} observatory
(Fabbiano \& Trinchieri 1987), and then observed with many other X-ray
missions. Its apparent X-ray luminosity varies between a few $10^{39}$ erg/s
and $3\times10^{40}$ erg/s in 0.3--10.0 keV (Feng \& Kaaret 2006; Mizuno et al.
2007), making it a typical ULX. The X-ray spectrum of NGC 1313 X-2 shows a
conventionally smooth and featureless profile, which can be modeled in
different ways (Stobbart et al. 2006). Many observations suggest that NGC1313
X-2 is a MXB system, with a BH mass $M_{\rm BH} \sim$ 50--100 $M_{\odot}$ and a
companion star $M_{\rm MS} \sim$ 8--20 $M_{\odot}$ (Zampieri et al. 2004;
Gris\'{e} et al. 2008; Liu et al. 2012). Zampieri et al. (2004) investigated
the X-ray properties with {\it Chandra}, {\it ROSAT}, {\it ASCA}, and {\it
XMM-Newton} data, and suggested that the mass of the compact remnant is $\sim$
100 $M_{\odot}$. The optical luminosity derived from {\it ESO} data indicated
that the mass of the companion is $\sim$ 15--20 $M_{\odot}$ (Zampieri et al.
2004). Gris\'{e} et al. (2008) analysed the {\it VLT/FORS1} and {\it HST/ACS}
photometric data, found that NGC1313 X-2 is associated with a star cluster with
an age of 20 Myr, and placed an upper mass limit of some 12 $M_{\odot}$ for the
companion. Using the constraints from the optical observations, Patruno \&
Zampieri (2008, 2010) computed evolution tracks of ULX counterparts on the
color-magnitude  diagram, and restricted NGC1313 X-2 to be a 50--100
$M_{\odot}$ BH accreting from a 12--15 $M_{\odot}$ main-sequence star. Liu et
al. (2009) firstly reported a possible orbital period $P \sim$ 6.12 days from
the source with {\it HST} observations. They also constrained the binary masses
and other parameters for NGC1313 X-2 by fitting the detected light curves and
radial velocity changes of the disk emission lines to an X-ray irradiated
binary model (Liu et al. 2012). However, the significance of this period was
low (Zampieri et al. 2012), and further spectroscopic monitoring observations
are required to firmly determine both the period and the mass function of
NGC1313 X-2. Of special interest is the surrounding ionized bubble nebula
($\sim$ 200 pc, Pakull \& Mirioni 2002), resembling ordinary supernova
remnants, but an order of magnitude larger. It is suggested that the continuous
outflow from the central X-ray source blows up the bubble (Pakull et al. 2006).

In this work, we attempt to use {\it XMM-Newton} archive data and the
information of ionized bubble nebula taken with {\it ESO VLT}, {\it HST}, and
the {\it SUBARU} telescope, to address the super-critical accretion flow in the
following questions. (1) How does the accretion disk evolve when the source is
in super-Eddington luminosity? (2) Can the source give the super-Eddington
emission for a long time? (3) Does the bubble nebula result from the outflow
formed in highly super-critical accretion? We describe the spectral and timing
analyses in next two sections, make the discussion in Section 4, and finally
present the conclusion in Section 5.

\section{Spectral Analysis}

We analyze all of the available {\it XMM-Newton} observations of NGC1313 X-2,
which were made prior to 2007. The data files from the EPIC camera are reduced
with the standard tools of XMM-SAS software version 11.0.0.. The light curve
above 10 keV is created, and then a count rate cut-off criterion is used to
exclude background flares. We select the data from good time intervals, by
setting FLAG = 0 and PATTERN $\leq$ 4 for PN data, and PATTERN $\leq$ 12 for
MOS data. Source spectra are extracted from circles with radius of 32\arcsec
and centered at the nominal position of NGC1313 X-2 (RA = 03:18:22.3, Dec =
-66:36:03.8, J2000), while background spectra from the same CCD chips as the
source and at a similar distance from the readout node. The spectral response
file are created using the SAS task {\it rmfgen} and {\it arfgen}. In several
cases, PN or MOS data are unavailable because the source was not covered by the
detector, or due to CCD gaps. When available, PN and MOS data from each
individual observation are fitted together for spectral analysis. The 0.3--10.0
keV band spectra are fitted with the HEAsoft X-ray spectral fitting package
XSPEC 12.7.1.  All spectra are rebinned to have at least 20 counts per bin to
enable the use of $\chi^2$ statistics.

We first fit the spectra with a powerlaw model, and a multicolor disk blackbody
model ({\it diskbb} in XSPEC) is added if it has a significance level above
99\% based on F-Test (Feng \& Kaaret 2006). All models in the paper also
include the interstellar absorption with the absorption column density being
treated as a free parameter. The unabsorbed total flux and the disk flux in
0.3--10.0 keV are calculated with the convolution model {\it cflux}. The
fitting results are shown in Table 1.

Unlike the cases in canonical low/hard and high/soft states in XRBs, ULXs
present no clear gaps, but a broad continuous distribution in photon index
($\Gamma$ $\sim$ 1.0-3.0, Feng \& Soria 2011). In NGC1313 X-2, the photon index
does not correlate with the disk flux, nor the total flux (Tables 1--3). In 8
of 14 observations, the source exhibits a significant disk component, and its
inner disk radius can be estimated from the {\it diskbb} model as:
\begin{equation}
R_{\rm disk}=\xi_{\rm cor} \mbox{ } N_{\rm disk}^{1/2} \mbox{ } \frac{D}{10
\mbox{ }{\rm kpc}} \mbox{ }\cos \theta ^{-1/2} \mbox{ } f_{\rm col}^{2}\mbox{ }
{\rm km},
\end{equation}
where $N_{\rm disk}$ is the normalization, the distance $D=4.13$ Mpc
(M\'{e}ndez et al. 2002), $\theta$ is the inclination angle of the disk,
$f_{\rm col}$ is the fractional change of the color temperature and $\xi_{\rm
cor}$ is the correction factor for the inner torque-free boundary condition
(Zhang et al. 1997). With the fitted inner disk radius $R_{\rm in}$ and its
temperature $T_{\rm in}$, the bolometric luminosity of the disk can also be
derived as: $L_{\rm disk}=4\pi R_{\rm in}^2\sigma_{\rm SB} T_{\rm in}^4$. The
Eddington luminosity is $L_{\mathrm{Edd}} = 1.3 \times 10^{38} \times
M/M_{\odot}$ erg s$^{-1}$ with $M$ being the mass of central compact object.
Assuming the BH mass is 50 ${M_{\odot}}$, and a set of reasonable parameters
(i.e., $\theta = 70\degr$, $\xi_{\rm cor}$ $=$ 0.41 (Kubota et al. 1998), and
$f_{\rm col}= 1.7$), we plot the ISCO-scaled radius of the inner accretion disk
versus the bolometric luminosity of the disk in units of $L_{\mathrm{Edd}}$ in
Figure 1(a), where the Galactic XRBs \footnote{In this paper, we use the term
``Galactic XRBs'' to include also LMC X-3.} data are also shown (see Paper I
and references therein).

It is interesting that the behavior of NGC1313 X-2 just resembles those of the
Galactic XRBs in the high-luminosity tail. However, in such cool disk model,
the disk contribution is overwhelmed by a powerlaw component, which may
possibly come from the Comptonization of blackbody seed photons (Yao et al.
2005). Using the simple {\it powerlaw} to depict the Compton component also
introduces another problem: it rises without limit at low energies, which
evidently disagrees with Comptonization. In this way, the flux of {\it
powerlaw} component is overestimated, and the flux of the soft thermal ({\it
diskbb}) component is suppressed, i.e. its normalization is reduced. That is,
the fits with {\it diskbb$+$powerlaw} give a cooler and smaller disk component
than reality (Yao et al. 2005; Steiner et al. 2009). Therefore, before deriving
the nature of the source, a physical Comptonization model is needed.

Here we use an empirical model of Comptonization, SIMPL ({\it simpl} in XSPEC),
to fit all 14 observations (Table 1 and Figure 1(b)). SIMPL offers a generic
approach to fitting Comptonized spectra with only two free parameters -- a
photon index and a scattered fraction of input seed photons. This model is
valid for a broad range of geometric configurations, and successfully matches
the behavior of other physical Comptonization models (see text and references
in Steiner et al. 2009). However, the fits fail to converge in 4 of 14
observations, and the errors of some parameters in those fits are unavailable
(Table 1). Compared with those cases in {\it powerlaw} model, the {\it diskbb}
component in {\it simpl} model contributes a larger fraction of X-ray emission
in 0.3--10.0 keV, even exceeds the non-thermal component (Tables 1--3). The
bolometric luminosity of disk component is up to several times of the
non-thermal emission in 0.3--10.0 keV in {\it simpl} model. However, we cannot
obtain reliable information of non-thermal emission, and can hardly extrapolate
its bolometric luminosity with the XMM-Newton's limited observing window
(0.3--10.0 keV). Bachetti et al. (2013) reported that NGC1313 X-2 is not
significantly detected by {\it NuSTAR} above 10 keV, suggesting that the
bolometric luminosity of NGC1313 X-2 is dominated by the disk component.

Because the disk component peaks where the interstellar absorption is of high
importance, the fitted disk flux is influenced by the absorption column density
$N_{\rm H}$ (Feng \& Kaaret 2007). Unfortunately, it is unclear whether the
variation of absorption column density in difference orbital phase is true. In
this way, we carry out a series of tests before obtaining the final results.

1. We replace the {\it wabs} model with another absorption model, e.g., the
Tuebingen-Boulder interstellar medium (ISM) absorption model ({\it tbabs} in
XSPEC), and the results of {\it tbabs*(diskbb$+$po)} are shown in Table 2.
However we fail to constrain some parameters with {\it tbabs*simpl*diskbb}
model, similar to the case with {\it wabs*simpl*diskbb}. The value of $N_{\rm
H}$ is insensitive to absorption models used, and the fitting results of {\it
tbabs} model are very similar to (or with slightly higher $N_{\rm H}$ than) the
results obtained with {\it wabs} model.

2. Feng \& Kaaret (2007) created two-dimensional plots of the 1 $\sigma$
confidence contours (model of {\it wabs*(diskbb$+$po)}), allowing both $N_{\rm
H}$ and $kT_{\rm disk}$ to vary (figure 1 in their paper). Taking the influence
of $N_{\rm H}$ into account, the anti-correlation between $L_{\rm disk}$ and
$T_{\rm disk}$ did not change but with larger scatter.

3. Assuming that $N_{\rm H}$ does not change among 14 observations, we fit all
observations simultaneously with the same value of $N_{\rm H}$, but leave the
other parameters untied (Table 3 and Figure 1(c)). In this way, the disk flux
of some observations are modified significantly; nevertheless, the data remain
in the sequence presented in the inner disk radius-luminosity plane, and our
discussion and conclusion are still valid.

Note that to obtain the accurate values of $R_{\rm disk}$ and $L_{\rm disk}$,
we need not only high quality X-ray data, but also the knowledge of $\xi_{\rm
cor}$, $f_{\rm col}$, $D$, and $\theta$. $R_{\rm disk}$ and $L_{\rm disk}$ are
related to these parameters: $R_{\rm disk} \propto \xi_{\rm cor} f_{\rm
col}^{2} D \cos \theta ^{-0.5}$ and $L_{\rm disk} \propto D^{2} \cos \theta$.
On the other hand, $R_{\rm ISCO}$ is inversely correlated with the spin
$a_{\ast}$ (Bardeen et al. 1972; Zhang et al. 1997). Unfortunately, we cannot
constrain these parameters well from any existing observation. With the same
parameters as those of XTE J1701-462, the NGC1313 X-2 disk evolution sequence
is parallel to and above the sequence of high-luminosity Galactic XRBs.
Changing the mass of BH decreases both $R_{\rm disk}/R_{\rm ISCO}$ and $L_{\rm
disk}/L_{\rm Edd}$, and thus the disk evolution sequence moves along the
diagonal direction rather than vertical direction in Figure 1. If the accretion
disk expands linearly with increasing luminosity from $\sim$ 0.3 $L_{\rm Edd}$,
and its evolution sequence is consistent with those of Galactic XRBs, we need
to pull down the sequence by increasing the inclination angle $\theta$, or
decreasing the value of $\xi_{\rm cor} f_{\rm col}^{2}$, and/or spin
$a_{\ast}$. However, it is suggested that the value of $\xi_{\rm cor} f_{\rm
col}^{2}$ might increase slowly with luminosity (e.g., Davis et al. 2005),
though the exact value in super-Eddington state is unknown. The large variation
amplitudes for the sinusoidal optical light curves indicated the possibility
that the inclination angle is larger than 70$\degr$ and smaller than 80$\degr$
(Liu et al. 2012). Because the value of $R_{\rm disk}/R_{\rm ISCO}$ decreases
with decreasing of spin $a_{\ast}$ and $L_{\rm disk}/L_{\rm Edd}$ is
independent of $a_{\ast}$, the disk evolution sequence moves down in the inner
disk radius-luminosity plane with a large retrograde spin. Thus, we adopt
$a_{\ast} = -1$ and $\theta = 80\degr$, and plot 10 of 14 convergent fittings
in the inner disk radius-luminosity plane (Figure 1(b)). In Figure 1, the error
of inner disk radius is derived from the error of disk component normalization.
On the other hand, we assume that the relative error of the disk bolometric
luminosity is equal to that of the 0.3-10.0 keV disk luminosity, which is
calculated with the convolution model {\it cflux}. All errors are in 90 \%
confidence level.

\begin{deluxetable}{llllp{1.0cm}llllll}
\tabletypesize{\tiny} \tablewidth{0pt} \tablecaption{BEST-FIT SPECTRAL
PARAMETERS OF NGC1313 X-2\label{tab:spec}} \tablehead{\colhead{Obs. Date} &
\colhead{Instruments} & \colhead{Exposure} & \colhead{$n_{\rm H}$} &
\colhead{$kT_{\rm disk}$} & \colhead{ $N_{\rm disk}$ }  & \colhead{$\Gamma$} &
\colhead{$flux_{\rm X}$} & \colhead{$flux_{\rm disk}$} &
\colhead{$\chi^2/$dof}} \startdata
&&&& wabs*(diskbb$+$po) &&&&\\
\noalign{\smallskip} \hline \noalign{\smallskip}

2000 Oct 17 &   PN          &   22.5            & $0.29_{-0.04}^{+0.07}$    & $0.21_{-0.05}^{+0.06}$    & $21.2_{-16.5}^{+82.2}$            & $2.21_{-0.15}^{+0.15}$  & $1.95_{-0.29}^{+0.24}$  &       $0.54_{-0.22}^{+0.47}$         &    104.1/95 \\
2003 Nov 25 &   PN/M1/M2    &   1.3/2.6/2.6     & $0.23_{-0.04}^{+0.05}$    & \nodata                   &       \nodata                     & $1.75_{-0.12}^{+0.12}$  & $3.16_{-0.23}^{+0.23}$  &       \nodata         &    101.7/88 \\
2003 Dec 21 &   PN/M1/M2    &   7.6/10.6/10.6   & $0.47_{-0.07}^{+0.10}$    & $0.11_{-0.01}^{+0.01}$    & $4066_{-1311}^{+12583}$           & $1.90_{-0.06}^{+0.09}$  & $10.00_{-1.29}^{+6.98}$ &       $5.58_{-3.15}^{+6.04}$         &    201.9/217 \\
2003 Dec 23 &   PN/M1/M2    &   3.3/8.9/9.2     & $0.26_{-0.02}^{+0.02}$    & \nodata                   &       \nodata                     & $1.70_{-0.05}^{+0.05}$  & $5.62_{-0.25}^{+0.27}$  &       \nodata         &    194.1/155 \\
2003 Dec 25 &   PN/M1/M2    &   7.0/9.0/9.1     & $0.26_{-0.02}^{+0.02}$    & \nodata                   &       \nodata                     & $2.19_{-0.07}^{+0.07}$  & $2.69_{-0.12}^{+0.13}$  &       \nodata         &    139.4/115 \\
2004 Jan 8  &   PN/M1/M2    &   9.2/12.9/13.0   & $0.28_{-0.02}^{+0.02}$    & \nodata                   &       \nodata                     & $2.36_{-0.07}^{+0.07}$  & $2.45_{-0.11}^{+0.12}$  &       \nodata         &    138.4/131 \\
2004 Jan 17 &   PN/M1/M2    &   5.3/8.4/8.4     & $0.27_{-0.03}^{+0.03}$    & \nodata                   &       \nodata                     & $2.38_{-0.10}^{+0.11}$  & $2.14_{-0.14}^{+0.21}$  &       \nodata         &    70.0/70   \\
2004 May 1  &   M1/M2       &   9.2/9.8         & $0.31_{-0.04}^{+0.04}$    & \nodata                   &       \nodata                     & $2.56_{-0.14}^{+0.15}$  & $2.19_{-0.24}^{+0.32}$  &       \nodata         &    74.6/84   \\
2004 Jun 5  &   PN/M1/M2    &   8.8/11.5/11.5   & $0.53_{-0.07}^{+0.08}$    & $0.12_{-0.01}^{+0.01}$    & $4858_{-3186}^{+9136}$            & $1.94_{-0.05}^{+0.06}$  & $12.02_{-2.25}^{+9.85}$ &       $6.96_{-3.60}^{+6.61}$         &    290.0/280 \\
2004 Aug 23 &   PN/M1/M2    &   10.0/14.7/14.9  & $0.22_{-0.05}^{+0.07}$    & $0.26_{-0.07}^{+0.10}$    & $5.78_{-4.3}^{+28.7}$             & $1.82_{-0.31}^{+0.22}$  & $1.51_{-0.10}^{+0.22}$  &       $0.42_{-0.11}^{+0.24}$         &    124.3/109 \\
2004 Nov 23 &   PN/M1/M2    &   12.4/15.5/15.5  & $0.28_{-0.04}^{+0.07}$    & $0.22_{-0.05}^{+0.07}$    & $14.4_{-11.2}^{+90.4}$            & $2.18_{-0.14}^{+0.14}$  & $1.95_{-0.29}^{+0.45}$  &       $0.45_{-0.19}^{+0.45}$         &    136.0/125 \\
2005 Feb 7  &   PN/M1/M2    &   9.6/12.8/13.2   & $0.56_{-0.07}^{+0.07}$    & $0.11_{-0.01}^{+0.01}$    & $12066_{-7203}^{+17429}$          & $1.89_{-0.05}^{+0.06}$  & $21.88_{-10.40}^{+3.83}$&       $10.46_{-4.91}^{+7.96}$        &    395.6/292  \\
2006 Mar 6  &   PN          &   17.4            & $0.53_{-0.09}^{+0.08}$    & $0.11_{-0.01}^{+0.01}$    & $6864_{-4878}^{+12033}$           & $1.99_{-0.07}^{+0.06}$  & $12.30_{-3.79}^{+6.75}$ &       $6.34_{-3.82}^{+6.45}$         &    241.3/195  \\
2006 Oct 15 &   PN          &   84.0            & $0.52_{-0.03}^{+0.05}$    & $0.11_{-0.01}^{+0.01}$    & $5726_{-1941}^{+5776}$            & $1.98_{-0.02}^{+0.04}$  & $12.30_{-1.83}^{+1.82}$ &       $7.08_{-2.15}^{+2.79}$         &    776.0/646  \\

\noalign{\smallskip} \hline \noalign{\smallskip}
&&&& wabs*simpl*diskbb &&&&\\
\noalign{\smallskip} \hline \noalign{\smallskip}

2000 Oct 17 &   PN        & 22.5                & $0.26_{-0.05}^{+0.08}$    & $0.21_{-0.05}^{+0.06}$    & $32.3_{       }^{+138.3}$     & $2.20_{-0.16}^{+0.14}$     & $1.70_{-0.29}^{+0.59}$    &       $0.81_{-0.19}^{+0.45}$         &     103.5/95  \\
2003 Nov 25 &   PN/M1/M2  & 1.3/2.6/2.6         & $0.60_{-0.18}^{+0.19}$    & $0.09_{-0.01}^{+0.02}$    & $45537_{      }^{+493264}$    & $2.08_{-0.19}^{+0.19}$     & $19.95_{-13.64}^{+59.48}$ &       $16.56_{-13.28}^{+60.71}$      &  90.6/86  \\
2003 Dec 21 &   PN/M1/M2  & 7.6/10.6/10.6       & $0.48_{-0.08}^{+0.09}$    & $0.11_{-0.01}^{+0.01}$    & $5196_{-3453}^{+11007}$       & $1.90_{-0.07}^{+0.08}$     & $11.22_{-3.63}^{+6.16}$   &       $6.18_{-3.32}^{+5.91}$         &    203.3/217 \\
2003 Dec 23 &   PN/M1/M2  & 3.3/8.9/9.2         & $0.42_{-0.14}^{+0.11}$    & $0.13_{-0.01}^{+0.02}$    & $1470_{-1225}^{+4851}$        & $1.80_{-0.09}^{+0.09}$     & $8.71_{-2.82}^{+5.74}$    &       $3.52_{-2.43}^{+5.31}$         &     190.0/153  \\
2003 Dec 25 &   PN/M1/M2  & 7.0/9.0/9.1         & $0.40_{-0.09}^{+0.11}$    & $0.12_{-0.02}^{+0.02}$    & $1355_{-1219}^{+5309}$        & $2.27_{-0.12}^{+0.12}$     & $4.57_{-1.94}^{+4.34}$    &       $2.33_{-1.64}^{+3.97}$         &     135.4/113  \\
2004 Jan 8  &   PN/M1/M2  & 9.2/12.9/13.0       & $0.28_{-0.02}^{+0.15}$    & $0.15_{-0.06}^{+0.10}$    & $133.2_{-41.2}^{+3037.7}$     & $2.32_{-0.09}^{+0.07}$     & $2.34_{-0.11}^{+2.44}$    &       $0.68_{-0.36}^{+2.04}$         &     135.9/129  \\
2004 Jan 17 &   PN/M1/M2  & 5.3/8.4/8.4         & $0.22_{-0.03}^{+0.11}$    & $0.25_{-0.12}^{+0.14}$    & $12.8_{     }^{+172.2}$       & $2.21_{-0.29}^{+0.13}$     & $1.74_{-0.26}^{+0.77}$    &       $0.72_{-0.09}^{+0.47}$         &     66.8/68  \\
2004 May 1  &   M1/M2     & 9.2/9.8             & $0.37_{-0.19}^{+0.29}$    & $0.12_{     }^{+0.43}$    & $774.3_{     }^{+31974.9}$    & $2.60_{-0.18}^{+0.30}$     & $2.75_{-0.98}^{+13.84}$   &       $1.22_{-1.21}^{+12.85}$        &    74.2/82 \\
2004 Jun 5  &   PN/M1/M2  & 8.8/11.5/11.5       & $0.53_{-0.08}^{+0.08}$    & $0.12_{-0.01}^{+0.01}$    & $5585_{-3520}^{+9232}$        & $1.94_{-0.06}^{+0.06}$     & $14.13_{-4.35}^{+6.77}$   &       $8.27_{-4.03}^{+6.55}$         &    290.7/280 \\
2004 Aug 23 &   PN/M1/M2  & 10.0/14.7/14.9      & $0.19_{-0.05}^{+0.08}$    & $0.27_{-0.08}^{+0.08}$    & $7.3_{-5.0}^{+36.7}$          & $1.80_{-0.27}^{+0.23}$     & $1.38_{-0.15}^{+0.36}$    &       $0.64_{-0.10}^{+0.21}$         &     124.4/109 \\
2004 Nov 23 &   PN/M1/M2  & 12.4/15.5/15.5      & $0.26_{-0.06}^{+0.07}$    & $0.21_{-0.04}^{+0.07}$    & $32_{-24.8}^{+143.6}$         & $2.18_{-0.16}^{+0.12}$     & $1.70_{-0.29}^{+0.66}$    &       $0.78_{-0.17}^{+0.49}$         &     135.0/125  \\
2005 Feb 7  &   PN/M1/M2  & 9.6/12.8/13.2       & $0.57_{-0.07}^{+0.07}$    & $0.11_{-0.01}^{+0.01}$    & $14015_{-8149}^{+18025}$      & $1.89_{-0.03}^{+0.06}$     & $17.78_{-5.48}^{+8.52}$   &       $11.95_{-5.19}^{+8.36}$        &    390.5/292 \\
2006 Mar 6  &   PN        & 17.4                & $0.52_{-0.08}^{+0.09}$    & $0.11_{-0.01}^{+0.01}$    & $7049_{-4459}^{+13354}$       & $1.99_{-0.06}^{+0.05}$     & $12.59_{-4.08}^{+6.91}$   &       $6.96_{-3.77}^{+6.64}$         &    241.4/195 \\
2006 Oct 15 &   PN        & 84.0                & $0.52_{-0.04}^{+0.04}$    & $0.11_{-0.01}^{+0.01}$    & $7326_{-2824}^{+4858}$        & $1.99_{-0.03}^{+0.03}$     & $12.88_{-2.41}^{+2.97}$   &       $7.77_{-2.18}^{+2.78}$         &    775.9/646  \\

\enddata

\tablecomments{Instruments: PN, MOS1(M1) or MOS2(M2); Exposure: clean exposure
in units of ksec for the corresponding instrument after background flares
excluded; $n_{\rm H}$: column density in units of $10^{22}$  cm$^{-2}$ along
the line of sight; $kT_{\rm disk}$: inner disk temperature (in units of keV) of
the multicolor disk component, which is unavailable when the disk component is
less than 99\% of confidence level; $N_{\rm disk}$: normalization of the disk
component; $\Gamma$: power-law photon index; $flux_{\rm X}$: 0.3--10 keV flux
in units of $10^{-12}$ erg cm$^{-2}$ s$^{-1}$; $flux_{\rm disk}$: 0.3--10 keV
disk flux in units of $10^{-12}$ erg cm$^{-2}$ s$^{-1}$; $\chi^2$/dof: $\chi^2$
and degree of freedom for the best-fit model; All errors are in 90\% confidence
level.}

\end{deluxetable}

\begin{figure}
\centering \epsscale{1.0} \plotone{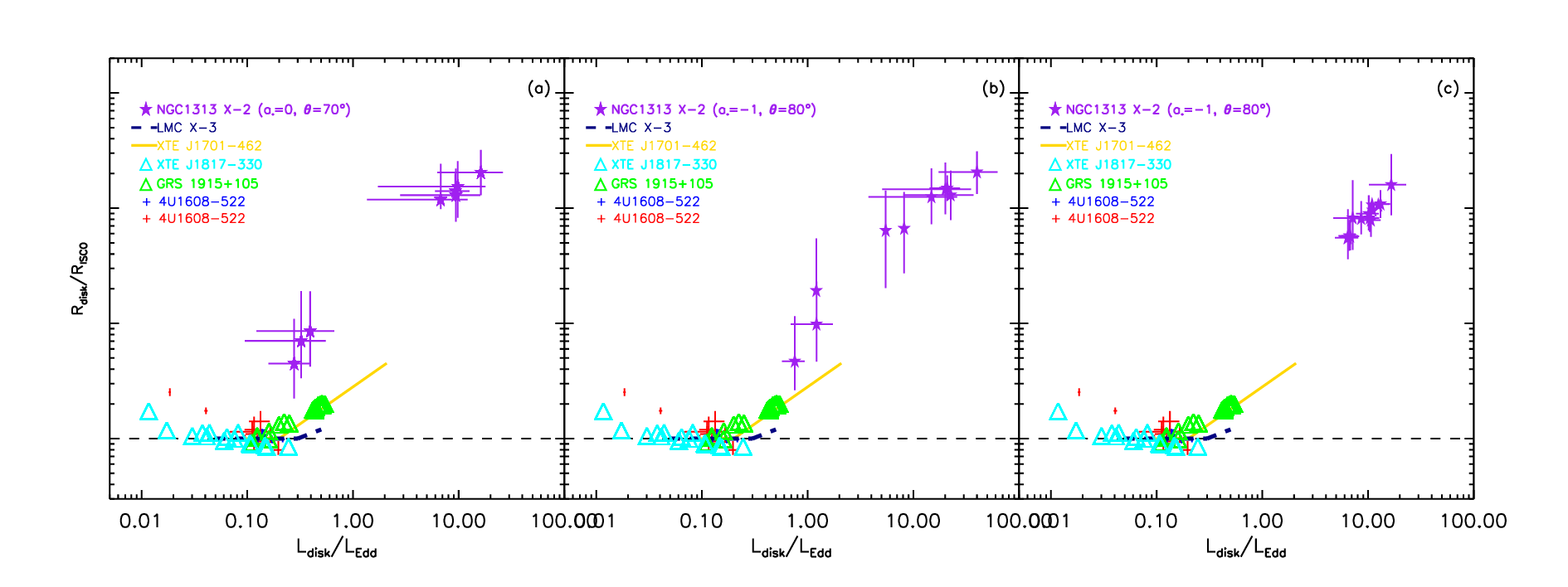} \caption{Relation between normalized
inner disk radius and normalized disk luminosity. NGC1313 X-2 data are shown
with the Galactic sources (Paper I) in the inner disk radius-luminosity plane.
a) $wabs(diskbb+powerlaw$) fit: A Schwarzschild BH with mass of $50 M_{\odot}$,
the color correction factor $f_{\rm col} = 1.7$, the correction factor
$\xi_{\rm cor}$ = 0.41, and an inclination angle $\theta = 70\degr$ are adopted
here. b) $wabs*simpl*diskbb$ fit: BH spin $a_{\ast} = -1$ and the inclination
angle $\theta = 80\degr$. c) the same as b), except that $N_{\rm H}$ is assumed
to be the same for all 14 observations. \label{fig1}}
\end{figure}

\begin{deluxetable}{llp{1.0cm}llllll}
\tabletypesize{\tiny} \tablewidth{0pt} \tablecaption{SPECTRAL FITS WITH THE
$tbabs*(diskbb+po)$ MODEL\label{tab:tabs}} \tablehead{\colhead{Obs. Date} &
\colhead{$n_{\rm H}$} & \colhead{$kT_{\rm disk}$} & \colhead{ $N_{\rm disk}$ }
& \colhead{$\Gamma$} & \colhead{$flux_{\rm X}$} & \colhead{$flux_{\rm disk}$} &
\colhead{$\chi^2/$dof}} \startdata

\noalign{\smallskip} \hline \noalign{\smallskip}

2000 Oct 17 &    $0.34_{-0.05}^{+0.09}$    & $0.19_{-0.04}^{+0.04}$    & $52.5_{-39.2}^{+279.9}$          & $2.23_{-0.14}^{+0.16}$  & $2.50_{-0.62}^{+0.37}$  &       $0.92_{-0.41}^{+0.94}$    &       100.8/95 \\
2003 Nov 25 &    $0.23_{-0.04}^{+0.05}$    & \nodata                   &       \nodata                    & $1.77_{-0.12}^{+0.13}$  & $3.22_{-0.25}^{+0.27}$  &          \nodata                &       100.0/88 \\
2003 Dec 21 &    $0.52_{-0.09}^{+0.09}$    & $0.12_{-0.01}^{+0.01}$    & $4150_{-3045}^{+8716}$           & $1.94_{-0.07}^{+0.07}$  & $10.08_{-2.22}^{+10.31}$&       $6.58_{-3.88}^{+7.04}$    &       199.9/217 \\
2003 Dec 23 &    $0.27_{-0.02}^{+0.02}$    & \nodata                   &       \nodata                    & $1.72_{-0.05}^{+0.05}$  & $5.70_{-0.21}^{+0.21}$  &          \nodata                &       192.4/155 \\
2003 Dec 25 &    $0.27_{-0.02}^{+0.02}$    & \nodata                   &       \nodata                    & $2.23_{-0.07}^{+0.07}$  & $2.76_{-0.14}^{+0.15}$  &          \nodata                &       140.8/115 \\
2004 Jan 8  &    $0.29_{-0.02}^{+0.02}$    & \nodata                   &       \nodata                    & $2.40_{-0.07}^{+0.07}$  & $2.54_{-0.14}^{+0.16}$  &          \nodata                &       142.4/131 \\
2004 Jan 17 &    $0.28_{-0.03}^{+0.03}$    & \nodata                   &       \nodata                    & $2.41_{-0.10}^{+0.11}$  & $2.24_{-0.18}^{+0.22}$  &          \nodata                &       71.5/70 \\
2004 May 1  &    $0.32_{-0.02}^{+0.05}$    & \nodata                   &       \nodata                    & $2.59_{-0.13}^{+0.18}$  & $2.31_{-0.28}^{+0.37}$  &          \nodata                &        75.2/84 \\
2004 Jun 5  &    $0.60_{-0.08}^{+0.08}$    & $0.12_{-0.01}^{+0.01}$    & $5790_{-3891}^{+9788}$           & $1.99_{-0.06}^{+0.06}$  & $16.45_{-4.40}^{+7.78}$ &       $10.73_{-5.52}^{+9.14}$   &       282.2/280 \\
2004 Aug 23 &    $0.24_{-0.06}^{+0.09}$    & $0.25_{-0.07}^{+0.08}$    & $8.3_{-6.1}^{+65.3}$             & $1.83_{-0.27}^{+0.24}$  & $1.61_{-0.25}^{+0.07}$  &       $0.53_{-0.16}^{+0.43}$    &       124.4/109 \\
2004 Nov 23 &    $0.33_{-0.05}^{+0.08}$    & $0.20_{-0.04}^{+0.05}$    & $35.0_{-27.4}^{+199.0}$          & $2.21_{-0.14}^{+0.14}$  & $2.37_{-0.56}^{+0.15}$  &       $0.74_{-0.32}^{+0.80}$    &       135.7/125 \\
2005 Feb 7  &    $0.61_{-0.08}^{+0.08}$    & $0.12_{-0.01}^{+0.01}$    & $10402_{-6470}^{+15678}$         & $1.92_{-0.06}^{+0.06}$  & $19.42_{-6.41}^{+6.50}$ &       $12.92_{-6.22}^{+10.24}$  &       392.7/292 \\
2006 Mar 6  &    $0.59_{-0.09}^{+0.09}$    & $0.12_{-0.01}^{+0.01}$    & $6035_{-4317}^{+12766}$          & $2.03_{-0.07}^{+0.07}$  & $15.86_{-5.62}^{+7.30}$ &       $9.31_{-5.46}^{+9.88}$    &       232.8/195 \\
2006 Oct 15 &    $0.58_{-0.04}^{+0.04}$    & $0.12_{-0.01}^{+0.01}$    & $6234_{-2720}^{+4293}$           & $2.03_{-0.03}^{+0.03}$  & $14.83_{-2.45}^{+2.24}$ &       $9.35_{-2.87}^{+3.72}$    &       759.2/646 \\
\enddata

\tablecomments{All errors are in 90\% confidence level.}

\end{deluxetable}

\begin{deluxetable}{llp{1.0cm}lllll}
\tabletypesize{\tiny} \tablewidth{0pt} \tablecaption{SPECTRAL FITS WITH THE
$wabs*simpl*diskbb$ MODEL\label{tab:combined}} \tablehead{\colhead{Obs. Date} &
\colhead{$n_{\rm H}$} & \colhead{$kT_{\rm disk}$} & \colhead{ $N_{\rm disk}$ }
& \colhead{$\Gamma$} & \colhead{$flux_{\rm X}$} & \colhead{$flux_{\rm disk}$}}
\startdata

\noalign{\smallskip} \hline \noalign{\smallskip}

2000 Oct 17 &    $0.45_{-0.02}^{+0.02}$    & $0.13_{-0.01}^{+0.01}$    & $1093_{-488 }^{+970 }$    & $2.41_{-0.11}^{+0.11}$    & $4.19_{-0.62}^{+0.76}$    & $2.93_{-0.55}^{+0.68}$ \\
2003 Nov 25 &    $0.45_{-0.02}^{+0.02}$    & $0.10_{-0.01}^{+0.02}$    & $8462_{-5972}^{+20181}$   & $1.94_{-0.11}^{+0.10}$    & $7.98_{-1.64}^{+2.24}$    & $4.72_{-1.51}^{+2.11}$ \\
2003 Dec 21 &    $0.45_{-0.02}^{+0.02}$    & $0.11_{-0.01}^{+0.01}$    & $3895_{-1667}^{+2878}$    & $1.89_{-0.04}^{+0.02}$    & $9.56_{-1.10}^{+1.32}$    & $4.72_{-1.01}^{+1.21}$ \\
2003 Dec 23 &    $0.45_{-0.02}^{+0.02}$    & $0.13_{-0.01}^{+0.01}$    & $2046_{-996 }^{+2061}$    & $1.82_{-0.05}^{+0.05}$    & $9.79_{-1.07}^{+1.30}$    & $4.48_{-0.94}^{+1.15}$ \\
2003 Dec 25 &    $0.45_{-0.02}^{+0.02}$    & $0.12_{-0.01}^{+0.01}$    & $2644_{-1280}^{+2854}$    & $2.31_{-0.08}^{+0.08}$    & $6.24_{-0.93}^{+1.15}$    & $3.86_{-0.83}^{+1.03}$ \\
2004 Jan 8  &    $0.45_{-0.02}^{+0.02}$    & $0.12_{-0.01}^{+0.01}$    & $2186_{-1029}^{+2178}$    & $2.46_{-0.08}^{+0.08}$    & $5.32_{-0.79}^{+0.97}$    & $3.29_{-0.70}^{+0.86}$ \\
2004 Jan 17 &    $0.45_{-0.02}^{+0.02}$    & $0.13_{-0.02}^{+0.02}$    & $1018_{-589 }^{+2162}$    & $2.43_{-0.14}^{+0.14}$    & $4.57_{-0.74}^{+1.01}$    & $2.92_{-0.61}^{+0.83}$ \\
2004 May 1  &    $0.45_{-0.02}^{+0.02}$    & $0.11_{-0.02}^{+0.03}$    & $2234_{-1609}^{+7844}$    & $2.67_{-0.15}^{+0.15}$    & $4.28_{-0.89}^{+1.21}$    & $2.53_{-0.73}^{+1.01}$ \\
2004 Jun 5  &    $0.45_{-0.02}^{+0.02}$    & $0.12_{-0.01}^{+0.01}$    & $2231_{-943 }^{+1599}$    & $1.89_{-0.04}^{+0.04}$    & $9.90_{-0.98}^{+1.16}$    & $4.18_{-0.88}^{+1.04}$ \\
2004 Aug 23 &    $0.45_{-0.02}^{+0.02}$    & $0.13_{-0.01}^{+0.01}$    & $1075_{-459 }^{+977 }$    & $2.23_{-0.11}^{+0.12}$    & $4.20_{-0.63}^{+0.77}$    & $3.05_{-0.57}^{+0.71}$ \\
2004 Nov 23 &    $0.45_{-0.02}^{+0.02}$    & $0.13_{-0.01}^{+0.01}$    & $1005_{-422 }^{+862 }$    & $2.38_{-0.09}^{+0.09}$    & $4.19_{-0.60}^{+0.73}$    & $2.87_{-0.54}^{+0.65}$ \\
2005 Feb 7  &    $0.45_{-0.02}^{+0.02}$    & $0.11_{-0.01}^{+0.01}$    & $3703_{-1491}^{+2428}$    & $1.81_{-0.03}^{+0.03}$    & $9.87_{-1.04}^{+1.22}$    & $4.50_{-0.95}^{+1.12}$ \\
2006 Mar 6  &    $0.45_{-0.02}^{+0.02}$    & $0.11_{-0.01}^{+0.01}$    & $3749_{-1844}^{+3824}$    & $1.94_{-0.04}^{+0.04}$    & $9.14_{-0.97}^{+1.17}$    & $3.72_{-0.87}^{+1.04}$ \\
2006 Oct 15 &    $0.45_{-0.02}^{+0.02}$    & $0.12_{-0.01}^{+0.01}$    & $3046_{-947 }^{+1316}$    & $1.94_{-0.02}^{+0.02}$    & $8.93_{-0.91}^{+1.06}$    & $4.10_{-0.84}^{+0.97}$ \\

\enddata
\tablecomments{All 14 observations are fitted simultaneously with the same
value of $N_{\rm H}$, and $\chi^2$/dof = 3043.4/2603 is obtained. All errors
are in 90\% confidence level.}

\end{deluxetable}

\begin{deluxetable}{llllllllll}
\tabletypesize{\tiny} \tablewidth{0pt} \tablecaption{SPECTRAL FITS WITH THE
$wabs*diskir$ MODEL\label{tab:diskir}} \tablehead{\colhead{Obs. Date} &
\colhead{$n_{\rm H}$} & \colhead{$kT_{\rm disk}$} & \colhead{$\Gamma$} &
\colhead{$kT_{\rm e}$} & \colhead{$Lc/Ld$} & \colhead{$rirr$} & \colhead{
$N_{\rm disk}$ }   & \colhead{$flux_{\rm X}$} & \colhead{$\chi^2/$dof}}
\startdata
\noalign{\smallskip} \hline \noalign{\smallskip}

2000 Oct 17 &    $0.23^{+0.03}_{-0.07}$    & $0.18^{+0.11}_{-0.03}$    & $2.20^{+0.12}_{-0.12}$    & $930.6^{   }_{-929.6}$    & $2.89^{     }_{-2.34}$    & $1.75^{     }_{-0.63}$    & $24.7^{+213.9}_{-20.5}$    & $1.53^{+0.58}_{-0.49}$  &  102.9/93 \\
2003 Nov 25 &    $0.17^{+0.12}_{-0.13}$    & $0.23^{+1.27}_{-0.07}$    & $2.06^{+0.86}_{-0.53}$    & $25.9^{     }_{-24.5}$    & $4.52^{     }_{-4.50}$    & $1.02^{+0.07}_{-0.01}$    & $14.0^{      }_{     }$    & $2.86^{+2.87}_{-0.87}$  & 89.6/84 \\
2003 Dec 21 &    $0.28^{+0.02}_{-0.09}$    & $0.11^{+0.15}_{-0.04}$    & $1.72^{+0.06}_{-0.05}$    & $1.83^{+0.16}_{-0.19}$    & $10.0^{     }_{-9.29}$    & $2.01^{     }_{-0.82}$    & $166.0^{+5310.3}_{-80.5}$  & $4.94^{+2.84}_{-0.48}$  & 181.6/215 \\
2003 Dec 23 &    $0.21^{+0.01}_{-0.01}$    & $0.08^{+0.01}_{-0.01}$    & $1.54^{+0.02}_{-0.02}$    & $1.41^{+0.01}_{-0.08}$    & $2.84^{+0.67}_{-0.53}$    & $1.00^{+0.01}_{-0.01}$    & $2249^{+2772 }_{-68  }$    & $4.81^{+0.98}_{-0.17}$  & 165.6/151 \\
2003 Dec 25 &    $0.39^{+0.11}_{-0.17}$    & $0.12^{+0.19}_{-0.03}$    & $2.14^{+0.18}_{-0.17}$    & $2.84^{     }_{-1.21}$    & $0.40^{     }_{-0.24}$    & $1.02^{     }_{-0.02}$    & $1503^{      }_{     }$    & $4.24^{     }_{     }$  & 134.1/111 \\
2004 Jan 8  &    $0.23^{+0.04}_{-0.12}$    & $0.16^{+0.22}_{-0.07}$    & $2.21^{+0.15}_{-0.21}$    & $3.13^{     }_{-1.60}$    & $8.12^{     }_{-4.30}$    & $3.27^{     }_{-1.50}$    & $15.8^{      }_{     }$    & $1.98^{+2.42}_{-0.47}$  &  135.3/127 \\
2004 Jan 17 &    $0.22^{+0.07}_{-0.05}$    & $0.31^{+0.08}_{-0.06}$    & $1.00^{+0.38}_{     }$    & $6.38^{     }_{-1.13}$    & $4.69^{+1.02}_{-2.02}$    & $1.02^{+0.02}_{-0.01}$    & $5.97^{+13.62}_{-4.37}$    & $1.92^{+0.41}_{-0.27}$  &  61.1/66 \\
2004 May 1  &    $0.22^{+0.02}_{-0.07}$    & $0.33^{+0.15}_{-0.15}$    & $7.34^{     }_{-6.21}$    & $1.00^{     }_{     }$    & $0.49^{+0.43}_{-0.10}$    & $1.00^{+0.01}_{-0.01}$    & $4.23^{+11.91}_{-4.23}$    & $1.32^{+0.29}_{-0.21}$  &  67.9/80 \\
2004 Jun 5  &    $0.23^{+0.05}_{-0.03}$    & $0.40^{+0.04}_{-0.04}$    & $2.78^{+0.53}_{-0.48}$    & $767.6^{   }_{-765.0}$    & $2.28^{+0.78}_{-0.59}$    & $1.01^{+0.01}_{-0.01}$    & $3.56^{+7.06 }_{-1.86}$    & $5.27^{+0.37}_{-0.27}$  &  230.6/278 \\
2004 Aug 23 &    $0.24^{+0.10}_{-0.10}$    & $0.22^{+0.10}_{-0.06}$    & $1.55^{+0.28}_{-0.28}$    & $8.84^{+9.86}_{-6.40}$    & $1.58^{+1.08}_{-0.81}$    & $1.01^{+0.06}_{-0.01}$    & $24.4^{      }_{     }$    & $1.63^{+0.79}_{-0.30}$  &   120.8/107 \\
2004 Nov 23 &    $0.22^{+0.07}_{-0.06}$    & $0.20^{+0.13}_{-0.12}$    & $2.17^{+0.10}_{-0.18}$    & $266.6^{   }_{-263.1}$    & $4.31^{     }_{-3.84}$    & $2.36^{+1.17}_{-1.22}$    & $11.5^{+240.6}_{-9.8 }$    & $1.45^{+1.00}_{-0.23}$  &  135.1/123 \\
2005 Feb 7  &    $0.23^{+0.07}_{-0.02}$    & $0.28^{+0.15}_{-0.08}$    & $1.61^{+0.32}_{-0.07}$    & $1.69^{+0.54}_{-0.22}$    & $2.60^{+0.91}_{-0.51}$    & $1.02^{+0.04}_{-0.01}$    & $11.6^{+32.8 }_{-10.0}$    & $4.89^{+0.78}_{-0.38}$  &  330.1/290 \\
2006 Mar 6  &    $0.25^{+0.01}_{-0.01}$    & $0.08^{+0.01}_{-0.01}$    & $1.59^{+0.02}_{-0.02}$    & $1.47^{+0.01}_{-0.08}$    & $2.30^{+0.36}_{-0.31}$    & $1.00^{+0.01}_{-0.01}$    & $2138^{+2141 }_{-33  }$    & $4.86^{+0.08}_{-0.07}$  &  178.8/193 \\
2006 Oct 15 &    $0.25^{+0.03}_{-0.02}$    & $0.27^{+0.07}_{-0.04}$    & $1.84^{+0.15}_{-0.13}$    & $2.12^{+0.52}_{-0.31}$    & $1.95^{+0.33}_{-0.17}$    & $1.01^{+0.01}_{-0.01}$    & $14.5^{+18.0 }_{-9.2 }$    & $4.54^{+0.33}_{-0.22}$  &  638.4/644 \\

\enddata

\tablecomments{$kT_{\rm disk}$: inner disk temperature (in units of keV);
$\Gamma$: power-law photon index; $kT_{\rm e}$: electron temperature; $Lc/Ld$:
ratio of luminosity in the Compton tail to that of the unilluminated disk;
$rirr$: radius of the Compton illuminated disk in terms of the inner disk
radius; $N_{\rm disk}$: normalization of the disk component; All errors are in
90\% confidence level.}

\end{deluxetable}

\section{Temporal Variability}

We extract the light curves from PN data with bin sizes of 200 s and 1 s from
the same regions as extracting the spectra. We use the task {\it epiclccorr} to
correct for good time intervals, dead time, and for background subtraction. We
first search for possible periodic modulations in its entire 0.2--10.0 keV
light curve, using the ftool {\it efsearch} with a 200 s bin-size and a period
resolution of 0.02 day. Figure 2 shows the result of chi-squared test over a
range of trial periods, i.e., $\chi^{2}$ versus period, in which the red arrow
marks the previously suggested orbital period of $P = 6.12$ days (Liu et al.
2009). There is no evidence of clear orbital modulation. This can be
interpreted as that the intrinsic X-ray spectral evolution and/or flux
variations dominate over the orbital modulation during the $\sim$ six years of
observation.

The power density spectrum (PSD) is constructed with the continuous exposure 1
s bin size light curves of 0.2--10.0 keV for each observation. The PSDs can be
fitted with a constant power density model indicating the Poisson noise level
for all but one observations. The PSD exhibits the red-noise behavior at low
frequencies ($\sim$ $10^{-3}$--$10^{-4}$ Hz) for the observation of 15 October
2006 (see figure 1 in Heil et al. 2009).

\begin{figure}
\centering \epsscale{0.8} \plotone{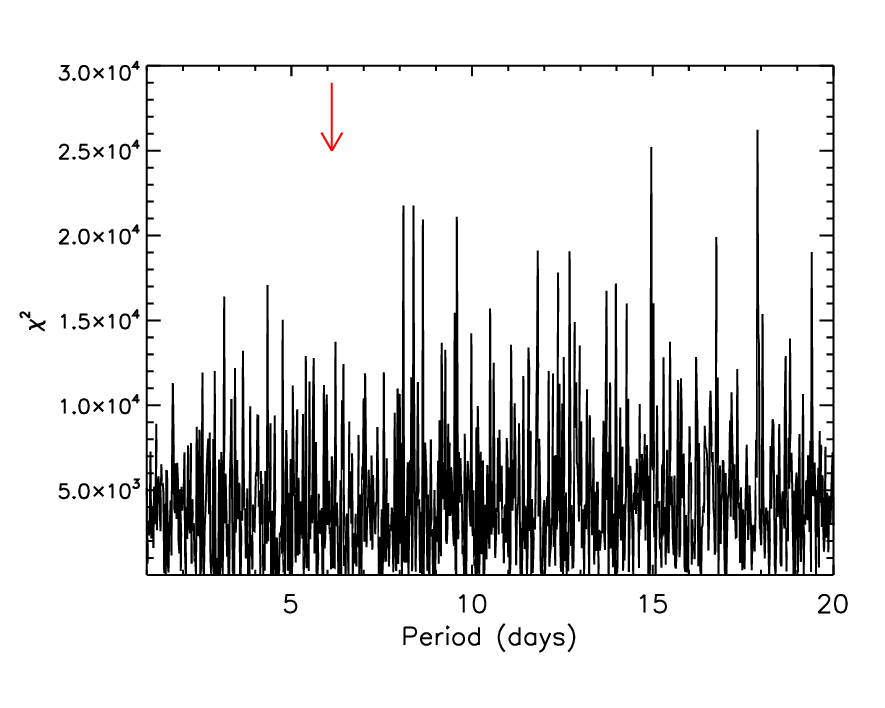} \caption{$\chi^{2}$ versus trial
period generated by {\it efsearch}, and the red arrow indicates the reported
orbital period of $P = 6.12$ days (Liu et al. 2009). \label{fig4}}
\end{figure}

NGC1313 X-2 was observed on 15 October 2006 with a long exposure of 123 ks.
After rejecting the data contaminated by the low amplitude flaring particle
background, we show the light curves of 0.2--0.8, 0.8--2.0, 2.0--10.0,
0.2--10.0 keV, and the hardness ratio between 0.2--0.8 keV and 2.0--10.0 keV in
Figure 3. The count rates of 0.2--10.0 and 2.0--10.0 keV change by a factor of
2 on timescales of $2\times10^{4}$ s, but the soft band emission is remarkably
steady. We calculate the fractional variability amplitude $F_{\rm{var}}$ with
200 s bin size light curves to quantify the variability,
\begin{equation}
F_{\rm{var}} = \sqrt{ \frac{S^{2} - \overline{\sigma^{2}}}{\bar{x}^{2}}},
\end{equation}
where $S^{2}$, $\sigma^{2}$, and $\bar{x}^{2}$ are the variance, mean error
squared, and mean count rate, respectively. The fractional variability
amplitude increases with energy, $F_{\rm{var}}$ = $4.8\pm7.2\%$, $5.5\pm1.7\%$,
and $10.2\pm1.7\%$ for 0.2--0.8, 0.8--2.0, and 2.0--10.0 keV, respectively.

\begin{figure}
\centering
\plotone{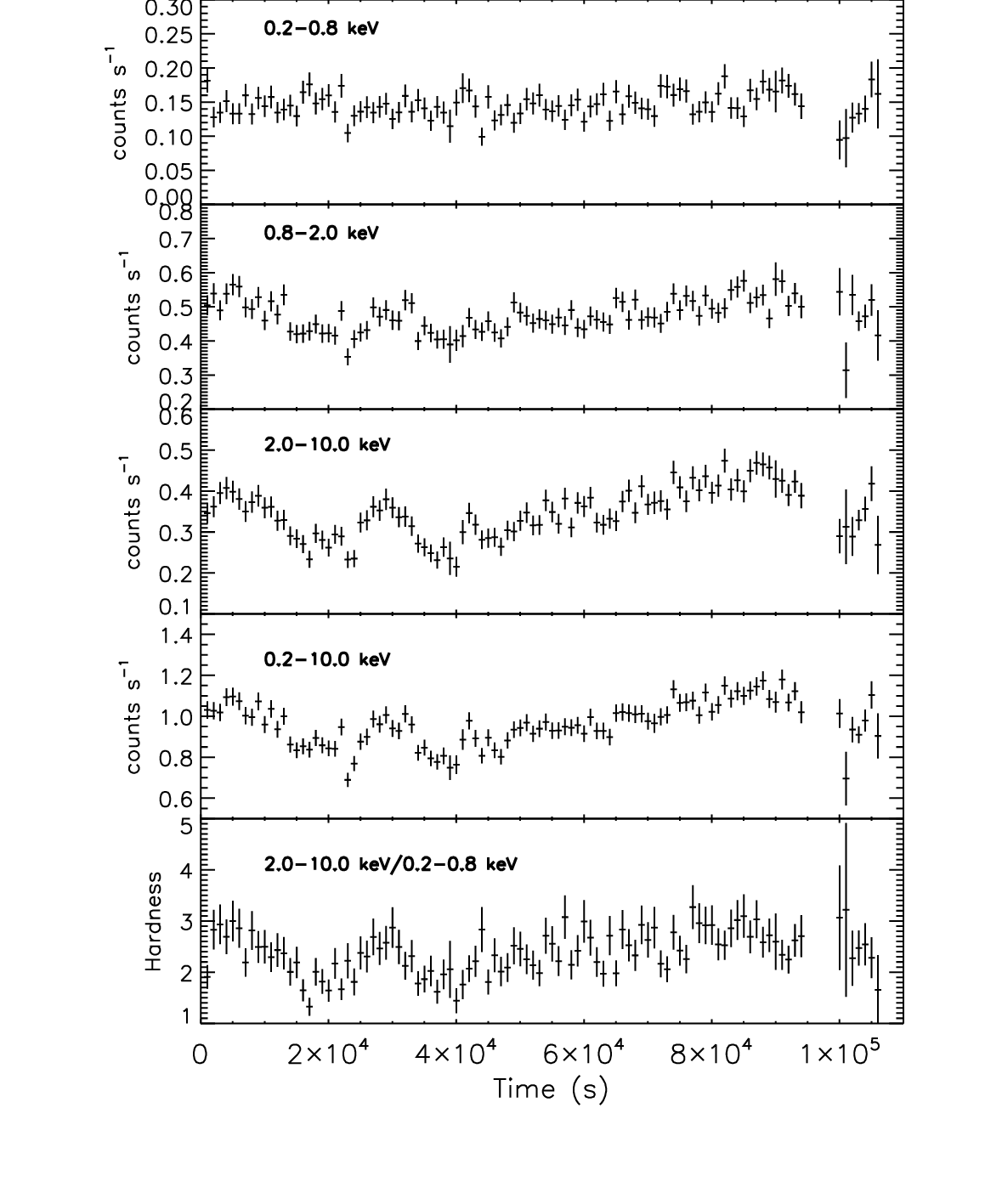} \caption{Panels from top to bottom show the 1000 s binned
background subtracted light curves in 0.2--0.8, 0.8--2.0, 2.0--10.0, 0.2--10.0
keV, and the temporal evolution of the 2.0--10.0 to 0.2--0.8 keV hardness
ratios, respectively.\label{fig5}}
\end{figure}

\section{Discussion}

\subsection{Truncated thin disk}
NGC1313 X-2 provides a textbook of ULX spectrum (Soria 2007; Stobbart et al.
2006). The cool disk component and the large inner disk radius detected in the
source have several interpretations (Feng \& Soria 2011). An obvious solution
is an IMBH accreting at sub-Eddington rates, and the accretion disk extending
to ISCO. However, Feng \& Kaaret (2007) found that the spectral evolution of
NGC1313 X-2 is inconsistent with $L_{\rm disk} \propto T_{\rm disk}^{4}$ law
expected for the standard accretion disk in IMBH. In addition, more and more
optical observations (e.g. the ESO VLT and HST data) suggest a MXB in the
NGC1313 X-2 (Gris\'{e} et al. 2008). Although the luminosity is model-dependent
for a given observation, it can hardly vary by an order of magnitude. Assuming
the mass of the BH is $\sim 50 M_{\odot}$ (Zampieri et al. 2004; Liu et al.
2012), the conservatively estimated X-ray luminosity $\sim 10^{40}$ erg/s
points to a super-critical accretion.

However, there is still no mature theory that can explain all X-ray properties
of super-critical accreting sources to date. At high luminosity, the standard
disk assumptions break down due to the effect of radiation pressure (Shakura \&
Sunyaev 1973). It is still unclear whether the accretion disk is geometrically
thin (Begelman 2002; Gu 2012), and whether it is truncated at a large distance
from the ISCO in super-Eddington luminosity (Soria 2007). McClintock et al.
(2006) argued that the inner disk edge is still located at ISCO, but its
emission is shaded by the outer disk when the luminosity goes higher and the
disk becomes thicker. Thus the fitting inner disk is not the true value but
just the larger radius at which the emission is not blocked by the outer disk.
The above effect of model depends on the viewing angle, and it is hard to
explain why the different sources give the same evolution pattern at high
luminosity. Besides, Gu (2012) investigated the vertical structure of accretion
disks in spherical coordinates and showed that the energy advection is
significant even for slightly sub-Eddington accretion disks, and therefore
argued that the non-negligible advection may help to understand why the
standard disk model is inaccurate above 0.3 Eddington luminosity.

Alternatively, Poutanen et al. (2007) suggested that the cool disk component
arose from the spherization radius of a super-critical disk. The model was
successfully used to explain several data of ULXs and the anti-correlation
between luminosity and temperature (Poutanen et al. 2007). They also predicted
that the spherization radius moves out and becomes cooler with increasing
luminosity. However, the observed trend is too steep for the predicted relation
of this model (Kajava \& Poutanen 2009). Moreover, the luminosity is
insensitive to the accretion rate when the source reaches and exceeds the
Eddington luminosity, $L \sim L_{\rm Edd}(1 + 0.6 \ln \dot{m})$ (where
$\dot{m}$ is the accretion rate in Eddington units). Thus, this model cannot
produce luminosity larger than 10 times of Eddington luminosity unless the
beaming effect is taken into account, and the radiative efficiency is less than
0.01 when the luminosity exceeds 3 times of Eddington luminosity (e.g. in
NGC1313 X-2).

Begelman (2002) put forward another plausible thin disk model that the
radiation pressure dominated accretion disk has strong density inhomogeneities
on scales much smaller than the disk scale height due to the development of
photon-bubble instabilities. The escaping flux from such disk could exceed
$L_{\rm Edd}$ by a factor of up to 10--100, depending on the BH mass.

In the specified bright states, the new-found NS XRB, XTE J1701-462, draws the
same track with BH XRB, GRS 1915+105, in the inner disk radius-luminosity plane
(Paper I; Neilsen et al. 2011). In contrast to the behavior of the inner disk
radius, the NS surface emission area maintained its small and nearly constant
size. All these information implied that the accretion disk is geometrically
thin, and moves out when the source is brighter than 0.3 $L_{\rm Edd}$ (see
Paper I for a detail discussion). Besides the X-ray spectral evolution, the
lack of short-time variability in the thermal component (0.2--0.8 keV band
emission, Figure 3) implies that the disk is truncated at a large distance from
the central BH. Investigating the {\it HST} archive data, Tao et al. (2011)
also suggested that the accretion disk in NGC1313 X-2 is geometrically thin,
and the optical emission was mainly due to its disk irradiation. Therefore, our
results are in favor of a truncated thin disk scenario.

In the low/hard spectral state, the thin disk recedes from ISCO, and is
irradiated by the Compton tail. This can change the inner disk temperature
structure from that expected from an unilluminated disk in the limit where the
ratio of luminosity in the tail to that in the disk, $L_{\rm c}/L_{\rm d} \gg
1$. The irradiated inner disk and Compton tail can also illuminate the rest of
the disk, and a fraction ($f_{\rm out}$) of the bolometric flux is thermalized
to the local blackbody temperature at each radius. This reprocessed flux
generally dominates the optical and UV bandpass of low-mass X-ray binaries
(Gierli{\'n}ski et al. 2008). Now the irradiated inner and outer disk model has
been developed to take account of these effects ({\it diskir} in XSPEC).

Assuming that the accretion disk in the super-Eddington state has the same
geometry as those in the low/hard state, we fit the spectra with {\it diskir}
model. Because of lacking of UV/optical data to cover the outer disk
information, we fix $f_{\rm out}$ (fraction of bolometric flux which is
thermalized in the outer disk) and {\it logrout} (the outer disk radius in
terms of the inner disk radius) to the default values. However the XMM-Newton
observing window (0.3--10.0 keV) is too limited to constrain the parameters of
{\it diskir} model (e.g., electron temperature and $L_{\rm c}/L_{\rm d}$, Table
4). Despite both $R_{\rm disk}$ and luminosity are systematically lower than
those in Table 3, the results indicate that $R_{\rm disk}$ is tens of ISCO, and
the luminosity is $\sim 10^{40}$ erg/s., which are still not in conflict with
our conclusion.

However we cannot obtain reliable information on the non-thermal emission, and
can hardly extrapolate its bolometric luminosity with the XMM-Newton limited
observing window (0.3--10.0 keV). The 0.3--10.0 keV disk flux is slightly
higher than the 0.3--10.0 keV flux of non-thermal emission for {\it
simpl*diskbb}, and the bolometric luminosity of disk emission is a few times of
the non-thermal emission 0.3--10.0 keV luminosity. As representative values, we
adopt $R_{\rm disk} \sim 50 R_{\rm ISCO}$ (or 200 $R_{\rm S}$, where $R_{\rm
S}$ is Schwarzschild radius), $L_{\rm total} \sim L_{\rm disk} \sim 10^{40}$
erg/s (Figure 1), and arrive at an accretion rate $\dot{M} \sim \frac{2 L_{\rm
total} R_{\rm disk}}{G M_{\rm BH}} \sim 1 \times 10^{-4}$ $M_{\odot}$/yr, and
the radiative efficiency $\eta_{\rm total} \sim \frac{G M_{\rm BH}}{R_{\rm
disk}}/c^{2} \sim 0.003$. Note that the radiative efficiency would be even
lower, and accretion rate would be even larger when taking the outflow into
account.

The response of the cold material (i.e. an accretion disk) to irradiation by
hard X-rays, may produce observable Fe emission lines, which have been
frequently detected in both AGNs and XRBs (Miller 2007; Gilfanov 2010). The
disk ionization state is dependent on the geometry of the accretion disk.
Compared with the thick disk, the thin disk would have higher number density
and thus lower ionization parameter. If the accretion disk is truncated at a
large radius, we could expect the Fe lines with Doppler shifts but no (or
small) gravitational red-shifts.

\subsection{Ultra-fast outflow}

There are two possible views on the way the accretion proceeds between $R_{\rm
disk}$ and $R_{\rm ISCO}$: 1, The accretion matter reaches the BH, but most of
the gravitational energy is released as the photons, which are trapped in the
flow and finally advected into the BH (e.g. Abramowicz et al. 1988); 2, Most of
gas is blown away by the radiation pressure, and the accretion rate decreases
with the disk radius (e.g. Begelman 2002; Gu \& Lu 2007). The most of power
released in the region between $R_{\rm disk} (\sim 50 R_{\rm ISCO})$ and
$R_{\rm ISCO}$ is eventually radiated as non-thermal power-law photons. In both
advection and outflow models, the radiative efficiency of the non-thermal
component in the inner region is significantly lower than the efficiency of
thermal emission in a standard disk (Soria 2007). In other words, $\eta_{\rm
total}/ \eta_{\rm disk} \sim L_{\rm total}/L_{\rm disk} < F$ (Soria 2007),
where $F = R_{\rm disk}/R_{\rm ISCO} \sim 50$ for NGC1313 X-2.

The emission of non-thermal component increases with the luminosity when the
source is close to its Eddington luminosity. The power-law emission of GRS
1915+105 and XTE J1701-462 in Figure 1 contribute less than 25\% of total X-ray
flux (McClintock et al. 2006; Paper I). Unlike the cases in canonical low/hard
and high/soft states in XRBs, ULXs present no clear gaps, but a broad
continuous distribution in photon index ($\Gamma \sim 1.0-3.0$, Feng \& Soria
2011). In NGC1313 X-2, the photon index does not correlate with the disk flux,
nor the total flux (Tables 1 and 2). On the whole the non-thermal component
contributes a larger fraction of accretion emission comparing with the bright
XRBs, but the disk emission still dominates over the non-thermal emission in
NGC 1313 X-2 (Section 4.1).

The strong radiatively driven outflows are produced in most highly
super-critical accretion models (e.g. Okuda et al. 2005; Begelman et al. 2006;
Ohsuga \& Mineshige 2011), and the power of outflows is approximately equal to
radiation. If the outflow is launched at the inner disk radius $\sim R_{\rm
disk}$ (or the transition radius), the speed of wind is possibly close to the
escape velocity, that is, mildly relativistic velocity from $R_{\rm disk} \sim
200\ R_{\rm S}$ (Figure 1). The ultra-fast outflows (UFOs) have been detected
in both AGNs and Galactic XRBs (Marshall et al. 2002; Crenshaw \& Kraemer 2012;
Tombesi et al. 2012). Recently, King et al. (2012) found the fastest disk wind
($v_{\rm wind} \sim 0.03$ c, c is the light speed) in the Galactic source, IGR
J17901-3624, with the help of the {\it Chandra} X-ray mission.

However, there is still no evidence of atomic iron features that could be
associated with massive outflowing material found in ULXs (Walton et al. 2012).
Note that the ionized nebulae are revealed not only in many other ULXs (Kaaret
et al. 2004; Abolmasov et al. 2007; Moon et al. 2011), but also a subset of
XRBs at very high mass accretion rates (Cooke et al. 2008; Pakull et al. 2010),
e.g. W50 nebula around SS433 (Fabrika 2004; Goodall et al. 2011). The
Doppler-shifted Fe lines together with many other emission lines originating
from the high temperature plasma in the conical jets were also detected in
SS433 (e.g., Marshall et al. 2002). However, there is no evidence of
relativistic jets found in NGC1313 X-2. In addition, the mechanism that UFOs
up-scatter the disk seed photons producing the observed non-thermal emissions,
has been intensively studied in literature (e.g. Kajava \& Poutanen 2009; Soria
2011). The observed variability shown in 2.0--10.0 keV light curve (Figure 3)
can be naturally ascribed to variations of UFOs.

The disappearance of UFOs can be interpreted as UFOs fully ionized or out of
our viewing angle. For a toy model, suppose that UFOs are launched from the
transition radius $R_{\rm disk}$ as a sphere expanding with the velocity
$v_{\rm UFOs}$, being the same order of magnitude as the escape velocity at
$R_{\rm disk}$, $v_{\rm UFOs} \sim \sqrt{2 G M / R_{\rm disk}}$. The gas
density at the radius of $R$ can be given by,
\begin{equation}
n_{\rm UFOs} \sim \frac{\dot{M}_{\rm UFOs}}{4\pi R^{2} v_{\rm UFOs} m_{\rm H}},
\end{equation}
where $m_{\rm H}$ is the mass of hydrogen atom (or proton) and the mass outflow
rate, $\dot{M}_{\rm UFOs} < \dot{M} \sim 10^{22}$ g/s. In this case, the
absorber ionisation parameter $\xi_{\rm ion}$ will be directly related to the
source luminosity $L_{\rm total}$ and $R_{\rm disk}$: $\xi_{\rm ion} >
\frac{L_{\rm total}}{n_{\rm UFOs} R^{2}}$. Plugging in the typical values,
$R_{\rm disk} \sim 200 R_{\rm S}$, $L_{\rm total} \sim 10^{40}$ erg/s (Figure
1), the expected ionization parameter $\xi_{\rm ion}$ is larger than $10^{4.7}$
ergs cm s$^{-1}$. That is, UFOs are fully ionized, a fact which accounts for
absence of the absorption line (Reynolds \& Nowak 2003).

However, the real geometry of UFOs could be more complicated than the above toy
model. If UFOs are clumpy, the localized gas density would be sufficiently high
to produce a weak and relatively broad Fe K absorption line (Neilsen et al.
2012), which can be tested with the next generation X-ray mission, e.g. {\it
Astro-H}. Another possibility is that the UFOs is anisotropic with small
covering factor, and the absorption feature appears only when UFOs occur in our
line of sight (Walton et al. 2012). In some rarely violent accretion systems,
the mass loss rate from outflows are much larger than the inflow rate derived
from the accretion disk (e.g. Pakull et al. 2010; Punsly 2011; Ponti et al.
2012). However, it is unlikely that the companions in ULXs can supply so much
accretion matter for a long time (next subsection). The power and low energy
conversion efficiency of radiatively driven outflows are probably close to
those of X-ray radiation ($\sim 0.003$) when sources are in super-Eddington
luminosity. That is, the total energy efficiency is less than 2 times of
radiative efficiency ($< 0.003 \times 2 \sim 0.006$).

\subsection{Short-lived ionized bubble nebula}
The observations of {\it ESO VLT}, {\it HST}, and {\it SUBARU} reveal that the
NGC1313 X-2 is embedded in a huge ionized nebula with a size of $\approx
200\times400$ pc, expanding at $v_{\rm bub}\approx 100$ km~s$^{-1}$, which is
the highly supersonic expansion speed (e.g. Ramsey et al. 2006). Pakull et al.
(2006) argued that the bubble is shock-excited with possible contribution due
to photoionization. If the bubble is inflated by a continuous outflow from the
central X-ray source, its age ($\tau_{\rm bub} \sim \frac{100\ {\rm pc}}{100\
{\rm km/s}} \sim 10^{6}$ yr) should be the same as the lifetime of the X-ray
active phase $\tau_{\rm acc}$,
\begin{equation}
\tau_{\rm acc} \sim \frac{M_{\rm acc}}{\dot{M}} \sim \frac{\eta m_{\rm acc}
c^{2}}{2 L_{\rm disk}},
\end{equation}
where $m_{\rm acc}$ is the total mass transferred from the companion to the BH
during the entire accretion lifetime, and it should be less than the mass of
the companion star, $M_{\rm acc} \sim 1-10 M_{\odot} < M_{\rm MS}$ (Zampieri et
al. 2004; Liu et al. 2012). Thus, a conservative estimate of the accretion age
is $\tau_{\rm acc} \sim 10^{4}-10^{5}$ yr. Taking the outflow into account, the
radiative efficiency would be even lower and $\tau_{\rm acc}$ even shorter. We
would like to emphasize that NGC1313 X-2 can emit $10^{40}$ erg/s as long as
$10^{6}$ yr only when it accretes 10 $M_{\odot}$ with a high radiative
efficiency ($\eta \sim 0.02$).

\subsubsection{Model I: Radiatively driven outflow}

We first investigate the possibility that NGC1313 X-2 is a transient with the
duty cycle of activity $\sim \frac{10^{4-5}\ {\rm yr}}{10^{6}\ {\rm yr}}$ and
the bubble nebula is resulted from the radiatively driven outflow during the
different X-ray active phases. In this scenario, only the earliest ($\sim
10^{6}$ yr ago) outflows are arriving at $\sim$ 100 pc from the central source
with $v_{\rm bub} \sim 100$ km~s$^{-1}$ and exciting the outmost part of the
bubble nebula; the later outflows have only reached shorter distances from the
central source and are ionizing the inner part of the bubble nebula. During the
propagation in the nebula, the purported outflows are decelerated because of
interaction with ISM. Here, we assume that the power of the radiatively driven
outflow is $\sim 10^{40}$ erg/s, which can be transferred 100\% from the
outflow to the bubble. Since the mechanical energy stored in the bubble nebula
is $E_{\rm bub} \sim 10^{52} \times (\frac{n}{1.0\ {\rm cm}^{-3}}) \times
{(\frac{R_{\rm bub}}{100\ {\rm pc}})^{3}} \times {(\frac{v_{\rm bub}}{100\ {\rm
km/s}})^{2}}$ erg, the source should be active no less than $10^{4}$ yr
($\tau_{\rm act} \geq$ $10^{4}$ yr) and accrete $\geq 1 M_{\odot}$ to inflate
the whole nebula for each duty cycle (or at least the earliest outburst), where
$n$ is the number density of the bubble nebula. Otherwise, the earliest
outflows should have been decelerated to a velocity lower than the observed
$\sim$100 km~s$^{-1}$, before reaching a distance of $\sim 100$ pc from the
central X-ray source. Therefore the source can only be active for a few cycles,
that is, the upper limit to the number of cycles is $\sim \frac{\tau_{\rm
acc}}{\tau_{\rm act}}$. The recombination cooling time scale is $\tau_{\rm
cool} \sim \frac{n}{\alpha_{\rm H} n^{2}} \sim \frac{1}{\alpha_{\rm H} n} \sim
10^{5} \times (\frac{1.0\ {\rm cm}^{-3}}{n})$ yr, where $\alpha_{\rm H} \sim
3\times10^{-13}\ {\rm cm}^{3}\ {\rm s}^{-1}$ is the recombination coefficient
(Cen \& Haimen 2000). It means that the observed ionized bubble nebula records
the last $\tau_{\rm cool} \sim 10^{5}$ yr ionized path of the outflows
generated in different active phases. In other words, the different parts of
the bubble nebula are resulted from the outflows produced in different epochs,
and the lower limit to the number of duty cycles is $\frac{R_{\rm bub}}{v_{\rm
bub} (\tau_{\rm cool}+\tau_{\rm act})}$. This model requires a high radiative
efficiency and a high energy transfer efficiency between the radiatively driven
outflow and ISM. The number density of ISM also plays a key role in this model:
The low-density ISM can be easily accelerated and has a long recombination
cooling time, or vice versa. Note that, the upper limit to the number of cycles
is close to (or even smaller than) the lower limit to the number of cycles for
this toy model with $n \sim 0.1-1.0\ {\rm cm}^{-3}$.

\subsubsection{Model II: Line-driven wind}

Here we investigate the possibility that the nebula is produced by the wind
driven by UV absorption lines. Because the accretion disk in NGC1313 X-2 is
relatively cool ($kT \sim 0.1 - 0.2$ keV), its radiation is dominated by
UV/soft X-ray emissions, which can be efficiently absorbed in the nebula. There
are plenty of absorption lines in UV spectral range, and their opacity is
several orders larger than the opacity for electron scattering. Thus, the
line-driven wind mechanism allows for efficient momentum transfer between
radiation and the nebula, resulting in fast moving plasma (Lucy \& Solomon
1970).

The size of ionized region can be calculated as the Str\"{o}mgren radius,
within which the rate of recombination balances the emission rate $\dot{N}_{\rm
ly}$ of ionizing photons from the central star. For typical O stars with
luminosities $L > 10^{4} L_{\odot}$, densities in Str\"{o}mgren sphere $n > 5
\times 10^{3}$ ${\rm cm}^{-3}$ and Str\"{o}mgren radii are $< 1$ pc (Habing \&
Israel 1979). The emission rate of NGC1313 X-2 estimated with XSEPC,
$\dot{N}_{\rm ly} \sim 10^{50-51} {\rm s}^{-1}$, is 1--2 orders larger than a
typical O star ($\sim 10^{49-50} {\rm s}^{-1}$, Martins et al. 2005). Moreover,
its surrounding ISM has a much lower number density $n \sim 0.1-1$ ${\rm
cm}^{-3}$, i.e., lower recombination rate. Thus, the UV photons from its
accretion disk can reach a large distance and form an ionized zone with a
radius of $R_{\rm S} \approx (\frac{3 \dot{N}_{\rm ly}}{4 \pi \alpha_{\rm H}
n^{2}})^{1/3} \approx 140 (\frac{\dot{N}_{\rm ly}}{10^{50}\ {\rm
s}^{-1}})^{1/3} (\frac{n}{1.0\ {\rm cm}^{-3}})^{-2/3}\ {\rm pc}$ (Cen \& Haimen
2000), which agrees with the size of the bubble nebula ($R_{\rm bub}$).

Adopting the spectral profile obtained in Section 2, we obtain that the
0.001--0.3 keV unabsorbed flux is $\sim$ 1-3 times of 0.3--10.0 keV unabsorbed
flux. Assuming solar abundances, the bubble nebula with number density $n \sim
0.1-1$ ${\rm cm}^{-3}$ and the radius $R_{\rm bub} \sim 100$ pc (Pakull \&
Mirioni 2002; Pakull et al. 2006), would absorb more than 50\% -- 90\% of
0.001--0.3 keV emission. The line-driven wind scenario also depends on the
absorption column density, the metal abundances of the bubble nebula, and the
cumulative radiation energy in UV/soft X-ray band, etc. The low metallicity of
the bubble nebula (e.g., Hadfield \& Crowther 2007; Zampieri \& Roberts 2009)
should decrease the momentum transfer efficiency between the radiation and the
interstellar medium. On the other hand, the UV emission would be enhanced
because of X-ray reprocessing at outer disk (Tao et al. 2011). Therefore it is
highly plausible that at least several percents of the total radiation energy
of NGC1313 X-2 is absorbed and stored in the bubble nebula.

If NGC1313 X-2 has been active for $\sim 10^{5}$ yr with $L \sim 10^{40}$
erg/s, only a few percents of cumulative radiation energy ($\sim 3 \times
10^{52}$ erg) transferred to the diffuse interstellar medium is sufficient to
fully ionize and power the expansion of the bubble nebula. We thus conclude
that this bubble nebula is very likely produced by photoionization and
line-driven expansion. This avoids the conflict between  the short accretion
age and the apparently much longer bubble age inferred from the expansion
velocity of the nebula, since the line-driven expansion happens almost
instantaneously everywhere in the nebula.

If NGC1313 X-2 is a transient, there are two possible cases of ionizing the
nebula in the line-driven wind scenario. {\it Case I}: If the radiation energy
in the last active cycle is larger than $E_{\rm bub}$, the bubble nebula is
created in the last active cycle with only $\sim$ 10\% of the energy required
by the radiatively driven outflow scenario. {\it Case II}: If the radiation
energy in the last active cycle is less than $E_{\rm bub}$ and the quench time
scale $\tau_{\rm quench}$ is less than the cooling time scale $\tau_{\rm
cool}$, the radiation energy input to the whole bubble nebula can be
accumulated during the previous active phases and the whole nebula is speeded
up gradually, since the radiation can reach the whole nebula almost
instantaneously. As a comparison, the radiatively driven outflow scenario needs
the input energy for each cycle to be larger than $E_{\rm bub}$ because the
input energy during the previous cycles cannot be accumulated in the whole
nebula, due to the limited shock propergation velocity (Section 4.3.1). Thus,
the line-driven wind scenario is preferred since the energy required by this
model can be much more easily satisfied in all possible scenarios.

\section{Conclusion}

In this study, we present both spectral and timing analyses of NGC1313 X-2 with
XMM-Newton archive data, together with the information of surrounding ionized
bubble, and make the conclusion as follows.

(1) The accretion disk of NGC1313 X-2 is truncated at a large distance $\sim
50\  R_{\rm ISCO}$, and draws the similar evolution pattern to luminous
Galactic XRBs. If the inner disk leaves the $R_{\rm ISCO}$ with increasing
luminosity from $\sim 0.3\ L_{\rm Edd}$, the fitting results imply that the
central BH has a fast retrograde spin.

(2) In the super-critical accretion, the radiative and mechanical energy
conversion efficiency is very low, that is, even lower than the radiative
efficiency of the nuclear fusion $\sim 0.007-0.009$. Therefore, the high
luminosity indicates that the companion star feeds the BH with an accretion
rate $\dot{M} \sim 1 \times 10^{-4}$ $M_{\odot}$/yr, which lasts less than
$10^{5}$ yr. We now are seeing an ephemeral feast in the source. If MXBs in the
early universe have the same short-lived, low energy conversion efficiency
accretion as NGC1313 X-2, they produce only limited feedbacks on their
environments.

(3) The UFOs accompanied with the super-critical accretion might be over
ionized and the absorption line might be hidden in super-Eddington X-ray
emission. If UFOs are clumpy or anisotropic, we can still expect weak and broad
iron line features with the extremely high energy resolution X-ray mission {\it
Astro-H} in near future.

(4) If the bubble is shock-excited and the velocity of the blastwave is equal
to the velocity inferred from the optical emission line ($\sim 100$
km~s$^{-1}$), the apparent bubble age would be up to $\sim 10^{6}$ yr (Pakull
\& Mirioni 2002; Pakull et al. 2006), that is much longer than the lifetime of
the X-ray active phase ($< 10^{5}$ yr). It may indicate that the source is a
transient with a duty cycle of activity of $\sim$ a few percent, and the bubble
nebula is resulted from the radiatively driven outflow formed in different
epochs. However, the upper limit to the number of cycles is close to (or even
smaller than) the lower limit to the number of cycles for this toy model with
$n \sim 0.1-1.0\ {\rm cm}^{-3}$.

(5)The Str\"{o}mgren radius of NGC 1313 X-2 agrees with the size of its bubble
nebula. Our calculations thus show that it is very likely that the bubble
nebula is photoionized and its expansion is line-driven, by the strong UV/soft
X-ray emission of the cool disk from NGC 1313 X-2, which requires much less
energy than that required by the radiatively driven outflow scenario and avoids
the conflict between the short accretion age and the apparently much longer
bubble age.

\acknowledgments{}  We thank the referee for  a thoughtful review. S.S.W.
thanks Weimin Gu, Wei Cui, and Yuan Liu for many valuable suggestions. This
work is partially supported with funding by the 973 Program of China under
grant 2009CB824800, by the National Natural Science Foundation of China under
grant Nos. 11133002, 10725313, 11003018, 11373036 and 11303022, and by the
Qianren start-up grant 292012312D1117210. S.S.W. is funded by the
T\"{U}B\.{I}TAK Co-funded Brain Circulation Scheme Co-Circulation program.

\end{document}